\documentclass[12pt]{article}
\usepackage{graphicx}
\usepackage{cite}
\def\pbnr{}
\def\speaker{Alexander Lenz}

\def\title{What did we learn in theory from the $\Delta A_{CP}$-saga? }
\def\affiliation{Institute for Particle Physics Phenomenology\\
Durham University, DH1 3LE, Durham, UK}

\newcommand\pubnumber{\pbnr}
\newcommand\pubdate{\today}
\def\Title#1{\begin{center} {\Large #1 } \end{center}}
\def\Author#1{\begin{center}{ \sc #1} \end{center}}

\newcommand{\OnBehalf}[1]{\sbox0{#1}\ifdim\wd0=0pt
        {}
	\else
	{\\on behalf of #1}
	\fi}
\newcommand{\SupportedBy}[1]{\sbox0{#1}\ifdim\wd0=0pt
        {}
	\else
	{\footnote{#1}}
	\fi}
\def\Address#1{\begin{center}{ \it #1} \end{center}}

\newcommand\pubblock{\includegraphics[width=5cm]{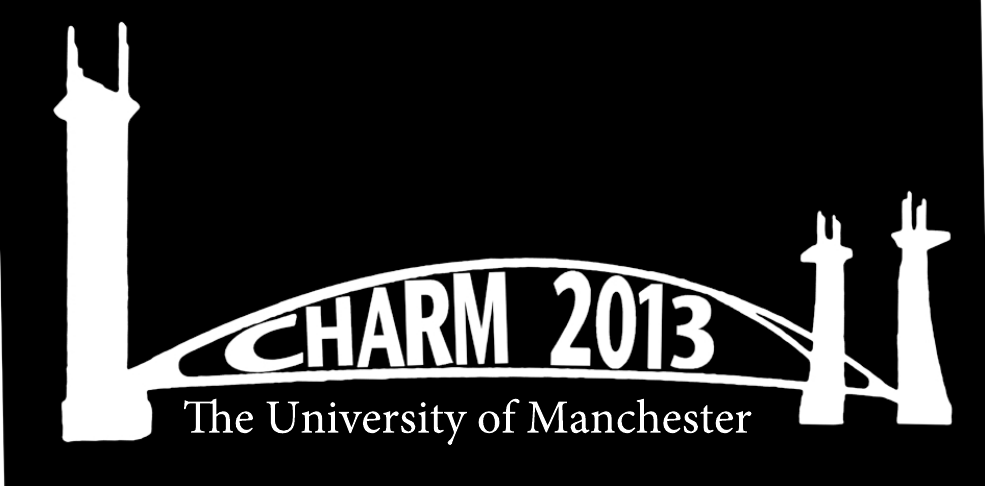}\hfill{\begin{tabular}{l} \pubnumber\\
         \pubdate  \end{tabular}}}
\newenvironment{Abstract}{\begin{quotation}  }{\end{quotation}}
\newenvironment{Presented}{\begin{quotation} \begin{center} 
             PRESENTED AT\end{center}\bigskip 
      \begin{center}\begin{large}}{\end{large}\end{center} \end{quotation}}
\def\Acknowledgements{\bigskip  \bigskip \begin{center} \begin{large}
             \bf ACKNOWLEDGEMENTS \end{large}\end{center}}
\def\venue{The 6$^{th}$ International Workshop on Charm Physics\\
(CHARM 2013)\\
Manchester, UK,  31 August -- 4 September, 2013}

\def\beq{\begin{equation}}
\def\eeq#1{\label{#1}\end{equation}}
\def\eeqn{\end{equation}}

\def\beqa{\begin{eqnarray}}
\def\eeqa#1{\label{#1}\end{eqnarray}}
\def\eeqan{\end{eqnarray}}

\let\bar=\overbar

\def\Dslash{\not{\hbox{\kern-4pt $D$}}}
\def\dslash{\not{\hbox{\kern-2pt $\del$}}}

\def\msb{{\bar{\ssstyle M \kern -1pt S}}}

\begin{document}
\begin{titlepage}
\pubblock

\vfill
\Title{\title}
\vfill
\Author{\speaker}
\Address{\affiliation}
\vfill
\begin{Abstract}
The measurement of CP violation in the charm sector triggered a lot of 
theoretical activities and re-considerations. Nevertheless currently
neither theory nor experiment have reached definite conclusions about
the origin of a large CP violating effect in hadronic $D$ decays.
We review briefly (part of) the current theory literature and present 
as the most important outcome of the many investigations 
initial steps in a long journey to understand the standard model contribution
to hadronic $D$ meson decays from first principles
as well as various 
control channels, that 
can be studied by experiment.
\end{Abstract}
\vfill
\begin{Presented}
\venue
\end{Presented}
\vfill
\end{titlepage}
\def\thefootnote{\fnsymbol{footnote}}
\setcounter{footnote}{0}
%

\section{Introduction}

The measurement of a non-vanishing CP-asymmetry in the charm sector
by the LCHb Collaboration \cite{Aaij:2011in} end of 2011 
- and also hints found by the CDF Collaboration \cite{Collaboration:2012qw} 
                    and the Belle Collaboration \cite{Ko:2012px} -
triggered an enormous amount of interest - about 200 citations at the time of writing this text - 
in the theory community, see e.g.
\cite{Bigi:2011em,Isidori:2011qw,Brod:2011re,Wang:2011uu,Rozanov:2011gj,Hochberg:2011ru,Pirtskhalava:2011va,Cheng:2012wr,Bhattacharya:2012ah,Giudice:2012qq,Altmannshofer:2012ur,Chen:2012am,Feldmann:2012js,Gedalia:2012pi,Mannel:2012qk,Li:2012cfa,Franco:2012ck,Brod:2012ud,Hiller:2012wf,Grossman:2012eb,Mannel:2012hb,Cheng:2012xb,Isidori:2012yx,KerenZur:2012fr,Bertolini:2012pu,Barbieri:2012bh,Chen:2012usa,Dolgov:2012ez,Delaunay:2012cz,Bhattacharya:2012kq,Botella:2012ju,DaRold:2012sz,Lyon:2012fk,Atwood:2012ac,Grossman:2012ry,Hiller:2012xm,Buccella:2013tya,Dighe:2013epa,Hiller:2013awa}.
Such a large amount of CP violation in the charm sector within the 
standard model was in contrast to common  text book wisdom, see e.g. 
\cite{Grossman:2006jg}\footnote{Although there 
were also previously to the $\Delta A_{CP}$ measurements
claims that CP violation of the order of $1\%$ cannot be completely excluded 
within the standard model, see e.g. \cite{Bobrowski:2010xg} for the 
case of $D$-mixing.}.
\\
Theorists unanimously identified NP to be the interpretation of this result,
but the theoretical literature still does not give a homogeneous picture,
whether NP abbreviates non-perturbative physics or new physics. Hence
the two main organisers of CHARM 2013 have asked me
to give an overview with the title 
{\it What did we learn in theory from the $\Delta A_{CP}$-saga?} - 
as a kind of independent point of view, since I did not publish on 
that topic\footnote{Scandalmonger claim it was because of a lack of 
valuable contributions to the Local Organisation Committee.}.
\\
The proposed title\footnote{To my knowledge the first use of the phrase 
``$\Delta A_{CP}$-saga'' was by Guy Wilkinson at BEAUTY 2013.}
contains the two keywords $\Delta A_{CP}$ and {\it saga},
on which I want to elaborate very briefly:
\begin{itemize}
\item $\Delta A_{CP}$ describes CP-violation in hadronic $D$-meson decays, see also
\cite{Kronfeld,Wilkinson,Jung,Schacht,Bertram,Neri,Fajfer,Schwartz,Charles,Hiller}. It
      is defined as the difference of the CP asymmetries in the $KK$ and 
      $\pi \pi$ final states.\footnote{See e.g. the reviews 
      \cite{Gersabeck:2012rp,Bediaga:2012py} for some experimental background on the 
      origin of this 
      definition.}
      \begin{equation}
      \Delta A_{CP} := A_{CP} (D^0 \to K^+K^-) - A_{CP}(D^0 \to \pi^+\pi^-) 
      \; .
      \end{equation}
      The first measurements 
      \cite{Aaij:2011in,Collaboration:2012qw,Ko:2012px} gave a combined value
      of (see e.g. \cite{Inguglia:2013aaa})
      \begin{equation}
      \Delta A_{CP} =  - 0.678 \pm 0.147 \% \, .
      \label{DeltaACPold}
      \end{equation}
      This large value seemed to be in clear contrast to previous
      expectations within the standard model, e.g. \cite{Grossman:2006jg}.
      Unfortunately the LHCb Collaboration performed further studies, 
      where the significance went down \cite{LHCupdate} or which even resulted in a 
      different sign \cite{Aaij:2013bra}. Taking these new numbers into
      account the new combination turns out to be \cite{Amhis:2012bh}
      \begin{equation}
      \Delta A_{CP} =  - 0.329 \pm 0.121 \% \, .
      \label{DeltaACPnew}
      \end{equation}
      The statistical significance for CP violation in hadronic $D$ decays
      went now down considerably, but the 
      central value is still larger than to be expected naively in the 
      standard model. Here clearly  further experimental input is 
      needed to settle this issue.
\item According to  Wikipedia {\it sagas} are described as  
      "... tales of worthy men,...", describing their ``battles and feuds''.
      Battles and feuds clearly took place between proponents of 
      large CP violating effects in $D$ decays within the 
      standard model and their opponents.
      Concerning the exclusion of women in the above quote, 
      I would like to contradict Wikipedia and mention that more 
      profound sources like \cite{Saga} emphasise the crucial - 
      albeit not always agreeable\footnote{Since I did not choose the title, I 
      refuse to investigate this point thoroughly.}
      -  role women played in sagas, e.g. {\it Hallgerd} from  Njal's Saga or 
      {\it Gudrun} in Laxd{\oe}la-Saga.
\end{itemize}
Starting the preparation of this talk, it was immediately clear that both the
experimental and theoretical literature still do not give a very unique picture.
Thus I started a poll  among (worthy) colleagues to get some idea about the opinions
inside the community and I posed the dictated title as the question:

\vspace{0.2cm}

\centerline{\it What did we learn from the $\Delta A_{CP}$-Saga?}

\vspace{0.2cm}

\noindent
I received the following instructive answers:
\begin{itemize}
\item ``...one should not jump on every { $3 \sigma$}-effect...''
\item ``...it is very difficult for me to say what happens in $D$-decays...''
\item ``...we need a deeper understanding of QCD...''
\item ``...{$\Delta A_{CP}$} in the standard model bigger than {$0.1\%$} is a stretch...
      ...original justification for considering enhanced penguins 
      in $D$ decays is somewhat weakened...''
\item ``...I don't really know what we can learn from { $\Delta A_{CP}$} 
      at this moment...'' 
\item ``...since { $\Delta A_{CP}$} seems now significantly smaller, 
      we believe that this is really a confirmation of our arguments for
      the standard model origin...''      
\item ``...penguins are still very large and currently one cannot not
      decide if this can be of standard model origin without additional assumptions...''
\item ``...{$\Delta A_{CP}$} should be at most a few times {$10^{-3}$} 
      in the standard model...''
\item ``...making (reliable) theoretical predictions about {$\Delta A_{CP}$} 
      is hard, but so is measuring it...''
\item ``...a value close to {$1\%$} is very unlikely in the standard model...''
\item ``...the most important thing I (re) learnt (over $\&$ over again)  
      is that {$3\sigma$} experimental results are not really reliable!
      But LHCb did a great job in focusing on $D$ decays''
\item ``...preliminary data can change a lot...
      we have to probe 3- and 4-body final states...''
\end{itemize}
To proceed from this uniformly inconclusive effort, let us next look closer
at the underlying structure of the $D$-meson decays.

\section{Singly Cabibbo suppressed D-meson decays}

$D$-meson decays into two hadrons can be classified by their dependence on 
the Cabibbo-Kobayashi-Maskawa (CKM) matrix elements:
\begin{enumerate}
\item \underline{CKM-favoured (CF) decays, like $D^0 \to K^- \pi^+$:}
      \\
      This decay proceeds via the quark level decay $c \to s u \bar{d}$
      and its CKM couplings are of order one. There are only tree-level
      contributions present.
\item \underline{Singly Cabibbo suppressed (SCS) $D$-meson decays, 
      like $D^0 \to \pi^- \pi^+$
      or $K^- K^+$:}
      \\
      This decay proceeds via the quark level decay  $c \to d u \bar{d}$
      (or $c \to s u \bar{s}$)
      and its CKM couplings are of the order of the Wolfenstein parameter 
      $\lambda$. Now both tree-level contributions and penguins are present.
\item \underline{Doubly Cabibbo suppressed (DCS) $D$-meson decays, 
      like $D^0 \to \pi^- K^+$:}
      \\
      This decay proceeds via the quark level decay $c \to d u \bar{s}$
      and its CKM couplings are of the order $\lambda^2$ and there are only 
      tree-level contributions present.
\end{enumerate}
To get some CP violating contributions we need at least two different 
amplitudes and in the standard model this can only be fulfilled by the SCS decays where we have 
both tree level amplitudes and penguin amplitudes.
Looking at the leading tree-level and penguin diagrams in the standard model
one finds the following general structure of SCS decays 
 \begin{equation}
      A (D^0 \to \pi^+ \pi^-) = 
                V_{cd}V_{ud}^* \left(A_{Tree} + A_{Peng.}^d \right)
              +
                V_{cs}V_{us}^* A_{Peng.}^s
              +
                V_{cb}V_{ub}^* A_{Peng.}^b \; .
      \label{amplitude1}
      \end{equation}
We have a tree-level amplitude $A_{Tree}$ with the CKM structure 
$ V_{cd}V_{ud}^*$ and three penguin contributions $A_{Peng.}^q$
with the internal quark $q=d,s,b$ and the CKM structure
$ V_{cq}V_{uq}^*$. All additional, more complicated contributions like e.g.
re-scattering effects can be put into the same scheme, see below. 
Already at that stage one can get a first feeling for the size
of CP violating effects in the SCS $D$ decays. Looking at the
arising CKM elements one finds:     
1)  {$V_{cd}V_{ud}^* \approx - V_{cs}V_{us}^*$} and both are almost real
and 2) {$ V_{cb}V_{ub}^*$} is very small.
Thus one expects very small CP violating effects in the standard model.
\\
To become a little more quantitative we look in more detail into
the structure of the decay amplitude and we use the effective 
Hamiltonian for the description of the decay, which is essential
for any numerical estimate.
The amplitude is given by
\begin{equation}
A \left( D^0 \longrightarrow \pi^+ \pi^- \right)
=
\langle D^0 | {\cal H}_{eff} | \pi^+ \pi^- \rangle \; ,
\end{equation}
with the effective Hamiltonian
\begin{equation}
 {\cal H}_{eff} = \frac{G_F}{\sqrt{2}}
\left[  \lambda_d \left( C_1 Q_1^d + C_2 Q_2^d \right)
      + \lambda_s \left( C_1 Q_1^s + C_2 Q_2^s \right)
      + \lambda_b \sum \limits_{i>3} C_i Q_i
\right] \;.
\end{equation}
The CKM structures are denoted by {$\lambda_x := V_{cx}^* V_{ux}$}.  
$Q_{1,2}^q$ are tree-level operators for the decays $c \to s + u \bar{s}$
(for $q=s$) and $c \to d + u \bar{d}$ (for $q=d$),
$Q_{3...}$ are the penguin operators triggering the decays
$c \to u + d \bar{d}$ and $c \to u + s \bar{s}$.
To obtain the amplitude one has to insert these operators in all
possible ways into Feynman diagrams of the effective theory with a 
$D^0$ as initial state and a pion pair in the final state.
One gets
\begin{eqnarray}
A &= &\frac{G_F}{\sqrt{2}}  \left[ 
  \lambda_d    \sum \limits_{i = 1,2} C_i  \langle Q_i^d \rangle^{T+P+E+R} 
+ \lambda_s    \sum \limits_{i = 1,2} C_i  \langle Q_i^s \rangle^{P+R} 
+ \lambda_b    \sum \limits_{i > 3} C_i  \langle Q_i^b \rangle^T 
\right] \; .
\label{amplitude2}
\end{eqnarray}
Here we use the following notation:
\begin{itemize}
\item {$\langle Q \rangle^T$}: tree-level insertion of the operator $Q$, 
\item {$\langle Q \rangle^E$}: insertion of the operator $Q$ in a weak exchange diagram, 
\item {$\langle Q \rangle^P$}: insertion of the operator $Q$ in a 
                                    penguin diagram,
\item {$\langle Q \rangle^R$}: insertion of the operator $Q$ in a re-scattering diagram.
\end{itemize}
As promised, Eq.(\ref{amplitude2}) still has the same principal structure
as Eq.(\ref{amplitude1}). To proceed further we have a closer look into 
the arising CKM elements. Expressing them in terms of three mixing angles
($\theta_{12}, \theta_{23}, \theta_{13}$) and one complex phase
($\delta_{13}$) one gets in the standard parameterisation of the quark mixing matrix
\begin{equation}
\begin{array}{ccclc}
\lambda_d & = & - s_{12} c_{12} c_{23} c_{13} & - c_{12}^2  \!\! \! & \!\! \! s_{23} s_{13} c_{13} { e^{i \delta_{13}}} \; ,
\\
\lambda_s & = & + s_{12} c_{12} c_{23} c_{13} & - s_{12}^2  \!\! \! &  \!\! \! s_{23} s_{13} c_{13} { e^{i \delta_{13}}} \; ,
\\
\lambda_b & = &                               & +         \!\! \!   &  \!\! \! s_{23} s_{13} c_{13} { e^{i \delta_{13}}} \; ,
\\
\end{array} 
\end{equation}
with the abbreviations $c_{ij} := \cos (\theta_{ij})$ and  $s_{ij} := \sin (\theta_{ij})$.
Keeping in mind the following hierarchies: 
$c_{ij} \propto {\cal O} (1)$,
$s_{12} \propto {\cal O} (\lambda)$,
$s_{23} \propto {\cal O} (\lambda^2)$,
$s_{13} \propto {\cal O} (\lambda^3)$,
one nicely sees that all potential CP violating effects are at least
$ {\cal O} (\lambda^5)$ and thus heavily suppressed in the standard model.
Using the unitarity of the CKM matrix - 
{$\lambda_s = - \lambda_d - \lambda_b$} - 
we get for the amplitude
{
\begin{eqnarray}
A \! \! & \! \!=  \! \!& \! \!\frac{G_F}{\sqrt{2}} \lambda_d 
\left[
{ 
 \sum \limits_{i = 1,2} C_i \langle Q_i^d \rangle^{T+P+E+R} 
\! \! - \! \! \sum \limits_{i = 1,2} C_i \langle Q_i^s \rangle^{P+R} 
}
\! \! +
\frac{\lambda_b}{\lambda_d} \left( 
{
  \sum \limits_{i > 3} C_i \langle Q_i^b \rangle^T 
\! \! - \! \! \sum \limits_{i = 1,2}    C_i \langle Q_i^s \rangle^{P+R} }
\right) 
\right] \; ,
\nonumber
\\
\label{amplitude3}
\end{eqnarray}
}
which can be abbreviated as
{
\begin{equation}
A  =: \frac{G_F}{\sqrt{2}} \lambda_d \; { T} 
      \left[ 1 + \frac{\lambda_b}{\lambda_d} { \frac{ P}{T}}\right] \; .
\label{amplitude4}
\end{equation}
}
The definitions of $T$ and $P$ can be simply read off from the comparison
of Eq.(\ref{amplitude3}) with Eq.(\ref{amplitude4}). Physical observables
like branching ratios or CP asymmetries can be expressed in terms of $|T|$, $|P/T|$ and the strong
phase $\phi = \arg( P/T)$ as
\begin{eqnarray}
Br &\propto & \frac{G_F^2}{2} |\lambda_d|^2 { |T|^2} \; ,
\\
 a_{CP} & = & 2 \left|\frac{\lambda_b}{\lambda_d}\right| \sin \delta_{13}
                 \left|\frac{ P}{ T}                \right| \sin \phi
              =  0.0012
                 \left|\frac{ P}{ T}                \right| \sin \phi \; .
\end{eqnarray}
In the last line we have used numerical input from \cite{Charles:2004jd} (see
\cite{Ciuchini:2000de} for similar results) for the CKM elements.
Up to this point all expressions are reliable and to an excellent accuracy 
exact and we are left with the subtlety 
that the values of $|P/T|$ and the strong phase $\phi$ are unknown.

\section{Welcome to the Sagaland}

The theoretical challenge in calculating the standard model value of 
$\Delta A_{CP}$ is now boiled down to the determination of matrix 
elements of the form $\langle D^0 | Q | \pi^+ \pi^- \rangle$.
Unfortunately we do not know if the tools that are very successful 
in $B$-decays or kaon decays can be applied to the charm system, whose 
mass scale is somehow
intermediate, not really heavy, but also not light.
\\
In order to proceed, additional assumptions have to be made, which might 
turn out in future to be unjustified.
Since time by time real battles and feuds were fought over these assumptions 
I decided to term them ideologies.
Roughly speaking one finds the following classes of assumptions:
\begin{itemize}
\item \underline{Ideology I:}   NP = non-perturbative physics
\item \underline{Ideology II:}  NP = New Physics
\item \underline{Ideology III:} Symmetry rules
\item \underline{Ideology IV:}  Experimentalists have to work harder
\end{itemize}
Unfortunately I will not be able to disproof any of these ideologies,
which are discussed in more detail below.

\subsection{This is clearly the standard model}
It is well known that non-perturbative effects can sometimes be huge, 
see e.g. the very famous $\Delta I = 1/2$ rule in $K \to \pi \pi$ decays.
Thus something similar might be acting in $\Delta A_{CP}$.
A good starting point for defending this ideology might be the assumption
of a order one strong phase. This gives
\begin{equation}
\left.
\begin{array}{lll}
\sin \phi          & = & 1 
\\
\Delta A_{CP}      & = & - 0.329 \%
\end{array}
\right\}
\Rightarrow
\left| \frac{P}{T}\right| = 1.3 \; .
\end{equation}
The current central experimental value for CP violation in hadronic $D$ decays 
could be reproduced 
if $|P/T|$ is enhanced roughly by one order of magnitude compared to naive standard 
model estimates, being discussed in the next subsection.
Such an enhancement might be compared to the famous {$\Delta I = 1/2$} rule,
which describes the ratio of the isospin zero part of the $K \to \pi \pi$
amplitude compared to the isospin two part:
\begin{equation}
            \frac{\Re \left( A_0 \right)}{\Re \left(A_2\right)} := 
            \frac{\Re \left[A (K \to \pi \pi)_{I=0}\right]}
                 {\Re \left[A (K \to \pi \pi)_{I=2}\right]} 
             = 22.5 \; .
            \end{equation}
This large experimental value contradicted naive estimates which expected a ratio
close to one. The inclusion of renormalisation group running effects enhanced the 
value to about two \cite{Gaillard:1974nj,Altarelli:1974exa} which was still far 
off the measured number.
\begin{equation}
            \frac{\Re \left(A_0\right)}{\Re \left(A_2 \right)} 
             = \left\{ 
              \begin{array}{cl}
                1 & \mbox{Naive}
                \\
                2 & \mbox{pert. QCD}
              \end{array}
              \right. \; .
            \end{equation}
One possible explanation for the large measured value might be a huge non-perturbative 
enhancement of penguin
contributions to the decay $K^0 \to \pi^+ \pi^-$. Such penguins would however only 
contribute to the $I=0$ final state.
If this would be the case, then it might seem kind of obvious to expect a similar, maybe less pronounced effect in
$D^0$ decays, that could then explain the large experimental value of $\Delta A_{CP}$. 
This potential analogy was mentioned several times in the theory literature
appearing after \cite{Aaij:2011in}, see e.g.\cite{Brod:2012ud,Pirtskhalava:2011va,Buccella:2013tya}
or \cite{Franco:2012ck,Atwood:2012ac,Hiller:2012xm};
earlier references 
are e.g. \cite{Abbott:1979fw} or \cite{Golden:1989qx}.
It is quite interesting to note at this stage that we recently gained quite some
considerable improvement in the theoretical understanding of the $\Delta I = 1/2$ rule.
Lattice calculations \cite{Boyle:2012ys} (based on earlier ideas of 
L{\"u}scher \cite{Luscher:1986pf,Luscher:1990ux,Luscher:1991cf}
and Lellouch, L{\"u}scher \cite{Lellouch:2000pv})
give a value that is now much closer to experiment
\begin{equation}
            \frac{\Re \left(A_0\right)}{\Re \left(A_2 \right)} 
             = 
              \begin{array}{cl}
                12.0    
              \end{array} \; .
            \end{equation}
Moreover the large value of \cite{Boyle:2012ys} arises from
a strong suppression of $A_2$ by ``surprising'' cancellations, which strongly 
depend on the values of quark masses, while no sizable penguin enhancement of
$A_0$ is found.
At first sight this might seem to discourage the idea that a large non-perturbative
enhancement is responsible for the measured value of  $\Delta A_{CP}$. But one has to keep
in mind, that it is currently not clear, how to relate the lattice results for 
the $K$-decays to $D$-decays. Compared to the kaon case we have in $D$-decays different
values of the quark masses (remember the strong dependence of the lattice result on that) 
and in a $D$-meson decay many more intermediate states are possible.
First steps for a determination of hadronic two-body $D$-decays on the lattice have been
presented in \cite{Hansen:2012tf}. So here we will have to wait for future theoretical
improvements, but the current status looks quite promising.
\\
Assuming $\sin \phi = 1$ the authors of \cite{Brod:2011re} got values like 
{$\Delta A_{CP} = 0.3 \%$} as ``plausible'' standard model predictions.
They pointed out that a large value of $|P/T|$ requires the matrix elements
of the $Q^s$ operators to be of similar size than the ones of the $Q^d$ operators -
this requirement can be read off from Eq.(\ref{amplitude3}). For such an 
enhancement of 
contracted contributions over uncontracted ones (and a corresponding breakdown of 
the $1/m_c$ expansion) the authors of \cite{Brod:2011re} were also
able to find - under
certain assumptions - both experimental as well as theoretical evidence. 

\subsection{This is clearly new physics}
It is well known that the heavy quark expansion (HQE) works very well in the $b$-system.
Recently it was even found that also quantities where this expansion might be 
questionable because of a small energy release in the decay, are very well described
by the theory, see e.g. the discussion in \cite{Lenz:2012mb}.
The prime example is the decay rate difference $\Delta \Gamma_s$ of neutral $B_s$ mesons.
This observable is governed by the quark level decay $b \to c \bar{c} s$, with a 
limited phase space in the final state because of the two massive charm quarks.
Recent measurements\cite{Aaij:2013oba,Aad:2012kba,Aaltonen:2012ie,Abazov:2011ry}
agree impressively well with the standard model prediction
\cite{Lenz:2011ti} based on the calculations in
\cite{Lenz:2006hd,Ciuchini:2003ww,Beneke:2003az,Beneke:1998sy,Beneke:1996gn}.
\begin{equation}
\left(\frac{\Delta \Gamma_s}{\Delta M_s} \right)^{\rm Exp}
\left/
\left(\frac{\Delta \Gamma_s}{\Delta M_s} \right)^{\rm SM}
\right.
= 0.92 \pm 0.12 \pm 0.20 \; ,
\end{equation}
where the first error is the experimental uncertainty and the second one the 
conservatively estimated theory error. The experimental average has been taken from 
\cite{Amhis:2012bh}.
\\
If the HQE works so well in the $b$-system it seems reasonable to
assume that this expansion delivers some rough but reliable 
estimates in the $c$-system. The analogue of 
$\Delta \Gamma_s$ in the neutral $D$-system is the mixing quantity $y$.  
Unfortunately there is an almost perfect GIM cancellation in the leading HQE 
term of $y$ and thus  this quantity is either governed by higher order terms in the HQE 
or by non-perturbative contribution, see e.g.\cite{Bobrowski:2010xg}. Thus a better
testing ground for inclusive $c$-decays seems to be the ratio of 
$D$-meson lifetimes, where no such cancellation occurs. Therefore one can test with these
observables if the experimental number agrees with the first few terms in
the HQE. 
Recent theoretical results \cite{Lenz:2013aua} seem to indicate
that the HQE is capable of explaining the large observed ratios, 
\begin{eqnarray}
              \frac{\tau (D^+)}{\tau D^0}^{\rm Exp} & = & 2.536 \pm 0.019 \; ,
              \\
              \frac{\tau (D^+)}{\tau D^0}^{\rm HQE} & = & 2.2 \pm 1.7(0.4) \; .
\end{eqnarray}
Currently the theory prediction is limited by the fact that there exists no
lattice calculation for the arising matrix elements of four quark 
operators. Such a calculation is clearly doable with current technology. 
Using vacuum insertion approximation and assuming large
uncertainties for the arising bag parameters one gets a theory uncertainty of 
$\pm 1.7$. Using similar errors as in the lattice results\footnote{Unfortunately the
most recent results \cite{Becirevic:2001fy} for $B$-meson lifetimes are from 2001!}
for $B$-meson lifetimes one gets a theory uncertainty of 
$\pm 0.4$. Hence doing the corresponding lattice calculation seems clearly to be a
worthy exercise. 
\\
Despite these being some interesting observations one has to keep in mind,
that it is a priori not clear what  inclusive arguments tell us 
about exclusive $D$-decays. Nevertheless we might gain or loose some confidence
in applying methods relying on a $1/m_c$ expansion, as we are doing now.
\\
``Naive'' perturbative estimates that seem to be valid in the $B$-system 
give, see e.g. \cite{Isidori:2011qw}, a very small expectation for $|P/T|$
 \begin{equation}
      \left| \frac{P}{T} \right| \approx \frac{\alpha_s}{\pi} \approx \frac{0.35}{\pi} 
           \approx 0.11 \; ,
       \end{equation}
which is of a similar size as values obtained within the framework of QCD factorisation 
for $D$-decays see e.g. \cite{Brod:2011re,Jung,Schacht,Hiller}
       \begin{equation}
      \left| \frac{P}{T} \right| \approx 0.08 ... 0.23 \; .
       \end{equation}
In contrast to the previous subsection we consider now 
$\sin \phi \approx 0.1$\footnote{I did not find any strong 
exclusion argument of this assumption, aimed to hyperbolise.} 
as a good starting point to obtain the standard model expectation for 
$\Delta A_{CP}$
\begin{equation}
         \Rightarrow a_{CP}^{\rm SM} \approx (1.0...2.8) \cdot 10^{-5} \; , 
\end{equation}
which is about two orders of magnitude lower than the measured number - thus clearly
new physics is at work.
\\
Now one could start to investigate whether the observed
value of $\Delta A_{CP}$ can be solely due to new physics.
This could be done model-independently, see e.g. 
\cite{Isidori:2011qw,Gedalia:2012pi,Grossman:2012eb,Isidori:2012yx,Barbieri:2012bh}
or by studying explicit models for physics beyond the standard model. To get some ideas
about the covered models I give a brief but incomplete list
of models and references:
a model with an extra chiral fermion generation \cite{Rozanov:2011gj,Feldmann:2012js,Dolgov:2012ez}\footnote{Adding
simply a fourth, perturbative, chiral family of fermions could recently be excluded
\cite{Djouadi:2012ae,Kuflik:2012ai,Eberhardt:2012gv} using electro-weak precision data and Higgs-searches.},
models with an extended scalar sector \cite{Hochberg:2011ru,Altmannshofer:2012ur},
SUSY \cite{Giudice:2012qq,Gedalia:2012pi,Hiller:2012wf},
models with an extended gauge sector \cite{Wang:2011uu,Altmannshofer:2012ur},
weakly couple new physics \cite{Mannel:2012hb},
composite Higgs models \cite{KerenZur:2012fr,DaRold:2012sz},
models with a L-R symmetry \cite{Bertolini:2012pu,Chen:2012usa},
models with extra dimensions \cite{Delaunay:2012cz,DaRold:2012sz},
models with extra vector quarks \cite{Botella:2012ju}
and
models with scalar diquarks \cite{Chen:2012am}.

\subsection{What did we learn from $SU(3)_F$?}

As long as we have no method to calculate the $D \to hh$ decays reliably
from first principles, one of the best strategies is probably the investigation
of symmetries, in particular flavour symmetries like $SU(3)_F$ and $U$-spin.
At first sight this approach seems to be discouraged by measurements like 
      \begin{equation}
       Br(D^0 \to K^+ K^-) = 2.8   \;   Br(D^0 \to \pi^+ \pi^-) \; ,
       \label{SU3B}
       \end{equation}
which are in sharp contradiction to the expectation of equal branching ratios
in a $SU(3)_F$-symmetric world.
A recent fit \cite{Hiller:2013awa}\footnote{This is an
update of \cite{Hiller:2012xm}, $SU(3)_F$ was also previously investigated e.g. in
\cite{Savage:1991wu,Chau:1991gx,Pirtskhalava:2011va,Bhattacharya:2012ah,Feldmann:2012js,Gedalia:2012pi,Li:2012cfa,Brod:2012ud,Grossman:2012eb,Cheng:2012xb,Grossman:2012ry,Franco:2012ck,Buccella:2013tya}.}
of the available $D \to PP$ decay data (16 branching ratios, 10 CP asymmetries and one strong phase)
including linear $SU(3)_F$ breaking terms in the theoretical expressions
finds that nominal 
$SU(3)_F$ breaking, i.e. about $30\%$ on the amplitude level can explain all data, 
also the one in Eq.(\ref{SU3B}). 
This conclusion can be found several times in the literature, nevertheless some comments 
are in order here:
\begin{itemize}
\item The fits in \cite{Hiller:2013awa} allow of course also large, i.e. non-nominal  
      $SU(3)_F$ breaking. 
\item Some authors e.g. \cite{Franco:2012ck,LUCA} come to different conclusions - i.e. $SU(3)_F$ breaking 
      has to be large.
\item If the $SU(3)_F$ breaking is nominal in the fits of \cite{Hiller:2013awa}, then the data
      requires a very strong penguin enhancement. This is driven not only by $\Delta A_{CP}$ but 
      also by CP asymmetries in $D^0 \to K_S K_S, D_s \to K_S \pi^+$ and $D_s \to K^+ \pi^0$.
\end{itemize}
To draw some definite conclusions about the actual size of the $SU(3)_F$ breaking and 
the penguin enhancement more precise data are necessary.

\subsection{Experimental cross checks}
Even if ``experimentalists have to work harder'' sounds a like theorists joke, 
following this path will probably be the most successful short-term way in understanding the
origin of $\Delta A_{CP}$.
In my opinion the most profound result of the numerous theory investigations 
\cite{Bigi:2011em,Isidori:2011qw,Brod:2011re,Wang:2011uu,Rozanov:2011gj,Hochberg:2011ru,Pirtskhalava:2011va,Cheng:2012wr,Bhattacharya:2012ah,Giudice:2012qq,Altmannshofer:2012ur,Chen:2012am,Feldmann:2012js,Gedalia:2012pi,Mannel:2012qk,Li:2012cfa,Franco:2012ck,Brod:2012ud,Hiller:2012wf,Grossman:2012eb,Mannel:2012hb,Cheng:2012xb,Isidori:2012yx,KerenZur:2012fr,Bertolini:2012pu,Barbieri:2012bh,Chen:2012usa,Dolgov:2012ez,Delaunay:2012cz,Bhattacharya:2012kq,Botella:2012ju,DaRold:2012sz,Lyon:2012fk,Atwood:2012ac,Grossman:2012ry,Hiller:2012xm,Buccella:2013tya,Dighe:2013epa,Hiller:2013awa}
was the identification of many decay channels for testing different theoretical assumptions.
I will give some examples below - to get an exhaustive overview the reader will have to 
study the above quoted literature:
\begin{itemize}
\item In the standard model one has only { $\Delta I =1/2$} penguins, thus any CP violation 
      in a { $\Delta I =3/2$} final state can only be due to new physics.
      An example for such a decay is, see e.g.
      \cite{Brod:2011re,Grossman:2012eb,Atwood:2012ac,Hiller:2012xm,Hiller:2013awa}
      \begin{equation}
      D^+ \to \pi^+ \pi^0 \; .
      \end{equation} 
\item The assumption that $SU(3)_F$ symmetry including nominal breaking holds, can be tested by 
      investigating decays like
      $D^0 \to K_S K_S$  (e.g. \cite{Brod:2012ud,Atwood:2012ac,Hiller:2012xm,Hiller:2013awa}), 
      $D_s \to K_S \pi^+$ (e.g. \cite{Hiller:2012xm,Hiller:2013awa}) 
      and $D_s \to K^+ \pi^0$ (e.g. \cite{Hiller:2012xm,Hiller:2013awa}). These are the decays that
      drive the penguin enhancement in the $SU(3)_F$ fits.
      Similar tests can be performed by comparing $D^0 \to K^+ K^-$ with $D^0 \to \pi^+ \pi^-$ or
       $D^+ \to \bar{K}^0 K^+$ with $D_s \to K^+ \pi^0$ (e.g. \cite{Hiller:2012xm,Hiller:2013awa}).
      In \cite{Grossman:2012eb,Grossman:2012ry} several sum rules have been set up, that can also
      be used to test the applicability of the $SU(3)_F$ symmetry.
\item Decays like $D^+ \to \phi \pi^+$ and $D_s \to \phi K^+$ are triggered by the same effective
      Hamiltonian as $\Delta A_{CP}$, so one might expect there also some sizable CP violating effects
      \cite{Feldmann:2012js}, if CP violation in hadronic $D$-decays is real.
\item A good new physics candidate for a penguin enhancement is the chromomagnetic operator, see
      e.g.
      \cite{Isidori:2011qw,Giudice:2012qq,Mannel:2012hb,Isidori:2012yx,Delaunay:2012cz,Lyon:2012fk}
      This possibility is, however, constrained by  D-mixing and direct CP violation in the 
      kaon system, $\epsilon'/ \epsilon$.
      Moreover it might lead to observale effects in e.g.
      $D \to P^+ P^- \gamma$, $ D \to \rho^0 \gamma$, $D \to \omega \gamma$ and 
      the corresponding CP asymmetries of these decays.
      Such a possibility could also lead to enhanced electric dipole moments of the neutron, see e.g.
      \cite{Mannel:2012hb}.
\item An investigation of multi-body $D$-decays was suggested e.g. in 
      \cite{Bigi:2011em,Grossman:2012eb}. First experimental results are discussed in
      \cite{Aaij:2013swa,Aaij:2013jxa,Harnew:2013mka}.
\item It is also instructive, see e.g. \cite{Dighe:2013epa} to measure the amount
      of indirect CP violation in the decays $D \to \pi^+ \pi^-$ and $D \to K^+ K^-$, denoted
      by $A_\Gamma(\pi \pi)$ and $A_\Gamma(K K)$. First results were just recently announced 
      at this conference
      and published in \cite{Aaij:2013ria}.
\end{itemize}


\section{Conclusion}

The LHCb results for CP violation in hadronic $D$-meson decays triggered a 
lot of interest. Unfortunately both the experimental situation as well as
the theoretical interpretation are still not settled.
So, ``what did we really learn in theory from the $\Delta A_{CP}$-saga?''
\\
The first result, although quite mundane and trivial: charm physics is interesting.
Of course, there was quite a number of researchers who found this out before,
but it seems that now charm physics came back on stage and it is now more generally
considered to be a hot topic, which leads to an increased number of theoretical and 
experimental activities.
\\
Next, some more profound achievements: even if we are still far away from a complete 
theoretical understanding of the hadronic structure of the $D \to hh$ decays within the
standard model, several important insights have been gained:
\begin{itemize}
\item Values for $\Delta A_{CP}$ of several per mille require a very sizable
      enhancement of penguin contributions. Some authors find such values within
      the standard model plausible, while one might also argue for standard model
      values of  $\Delta A_{CP}$ of the order of $10^{-5}$.
\item New lattice results for kaon decays do not find a huge non-perturbative
      enhancement of penguin contributions in $K \to \pi \pi$ decays.
\item It is a priori not clear how to extrapolate the lattice results for the kaons 
      to the charm sector. But first investigations of hadronic $D$-meson decays on the 
      lattice seem to indicate that such a endeavour might be doable, even if this will 
      be a long term project.
\item Recently we also learnt that the expansions in the inverse of the heavy quark mass work
      also very well for ``critical'' inclusive observables in the $b$-quark system. Thus 
      it might seem not completely hopeless to apply such an expansion in the charm system.
      The ideal testing ground for this would be the ratios of $D$-meson lifetimes.
      To make some decisive statements, the 
      lattice results for the arising matrix elements of four-quark 
      operators - the ``only'' missing part - have to be determined. A task which clearly can
      be done with present techniques. This test will be very instructive, but even if it 
      turns out that $D$-meson lifetime ratios can be described within the heavy quark 
      expansion, it is still not completely clear what this means for exclusive decays.
      Nevertheless it might strengthen or weaken our confidence in applying methods like 
      QCD factorisation to $D \to hh$ decays.
\end{itemize} 
The third class of insights were gained from $SU(3)_F$ symmetry investigations. As long as we cannot 
treat the problem from first principles, symmetries are probably the best tools to approximate the
problem. Currently there are many indications that nominal $SU(3)_F$ breaking combined with
a large enhancement of penguins could describe all $D \to PP$ data. To draw definite conclusions, however, 
more data are necessary.
\\
Finally the most concrete outcome of the theoretical investigations so far was the identification
of numerous experimental channels, that allow a definite test of different theoretical assumptions.
\\
Despite there were several interesting theory results obtained by investigations of new
physics effects in $\Delta A_{CP}$ I did not concentrate on individual models, because I consider
the question whether the results in Eq.(\ref{DeltaACPold}) or Eq.(\ref{DeltaACPnew})
can be of standard model origin as pivotal. 
\\
All in all, quite a lot has been already learnt in theory from the $\Delta A_{CP}$-saga
and it seems the story will go on, both in theory and experiment. 
Shedding more light on the possible size of penguin contributions will also
be crucial in $B$-physics, e.g. the penguin pollution in $B_s \to J / \psi \phi$.
\\
A final comment: since the current literature is still (necessarily) full of prejudices 
it might also be interesting to know the prejudice of the author of this proceedings:
the old central value in Eq.(\ref{DeltaACPold}) is a clear signal for new physics
and even if it is a kind of hair-splitting: going back to the original meaning
of plausible (stemming from {\it lat. plausibilis}, which again originates 
from {\it plaudere}), I think that a standard model origin of a large value for
{ $\Delta A_{CP}$} 
is not plausible, but currently it can also not be excluded completely.
%
%

\Acknowledgements
I am grateful to I. Bigi, J. Brod, T. Feldmann, G. Hiller, G. Isidori, A. Kagan, 
J. Kamenik, H. Li, U. Nierste, L. Silvestrini, A. Soni and J. Zupan
for taking part in the poll. I also benefited a lot from discussions with M. Jung, J. Kamenik, 
A. Kronfeld, S. Schacht and L. Silvestrini. 
Finally many thanks to Marion Lenz, Mathilde Lenz, Rainer Schenk and Ursula 
Schenk for providing me the time for writing up these proceedings in a period, where I should
have spent my time differently.

\end{document}